\documentclass{svproc}
\usepackage{url}

\usepackage{graphicx}   
\usepackage{float}    
\usepackage{multicol}        
\usepackage[bottom]{footmisc}
\usepackage{lineno}
\usepackage{sidecap}

\usepackage{blindtext}

\begin{document}
\mainmatter              
\title{Recent bottomonium measurements in pp, p--Pb and Pb--Pb collisions at forward rapidity with ALICE at the LHC}
\titlerunning{Bottomonium at forward rapidity with ALICE }  
%
\author{Wadut Shaikh${^{1,}}$${^2}$ (for the ALICE Collaboration)}
\institute{ $^1$Saha Institute of Nuclear Physics, Kolkata-700064, India\\
$^2$Mugberia Gangadhar Mahavidyalaya, Purba Medinipur-721425, India\\
\email{wadut.shaikh@cern.ch}
}

\maketitle              
 \vspace{-0.3cm}
\begin{abstract}

In ultrarelativistic nucleus-nucleus collisions, a deconfined state of strongly interacting matter is thought to be produced, commonly known as the quark--gluon plasma (QGP). Quarkonia, bound states of a heavy quark and antiquark, are an important probe to study the properties of the QGP. At the LHC, bottomonium (b$\rm\overline{b}$) is of particular interest to study the QGP complementarily to the lighter charmonium (c$\rm\overline{c}$) system.  In addition to the measurements in nucleus-nucleus collisions, reference measurements in proton-proton and proton-nucleus collisions have been also carried out in order to better understand the bottom quark production and cold nuclear matter effects. ALICE measures the bottomonium production in the dimuon decay channel at forward rapidity ($2.5<y_{\rm {lab}}<4.0$) with the muon spectrometer. In this contribution, the recent measurements of bottomonium nuclear modification factors and azimuthal anisotropies in Pb--Pb collisions are presented. The bottomonium production in p--Pb and pp collisions are also discussed.
\keywords{QGP, Quarkonium, CNM, Nuclear modification factor}
\end{abstract}
 \vspace{-1.0cm}
\section{Introduction}
 \vspace{-0.1cm}
Quarkonia are useful probes to investigate the properties of the deconfined medium created in ultra-relativistic heavy-ion collisions. The modification of the ground charmonium state (J/$\psi$) production at LHC energies in heavy-ion collisions with respect to the binary-scaled yield in pp collisions has been explained as an interplay of the suppression~\cite{Matsui} and the regeneration mechanisms~\cite{regenaration1,regenaration2}. Bottomonia ($\Upsilon$) are also expected to be suppressed inside the QGP by the color-screening effect and medium-induced dissociation~\cite{Rothkopf:2019ipj}. For the $\Upsilon$ family, the regeneration effects are expected to be negligible due to the smaller number of b quarks produced in the collisions~\cite{Grandchamp:2005yw}. In addition, the regenerated quarkonia are expected to inherit the azimuthal anisotropy that the constituting quarks may have obtained by participating to the collective motion in the QGP. However, the Cold Nuclear Matter (CNM) effects which include shadowing, parton energy loss, interaction with hadronic degrees of freedom may also lead to a modification of bottomonium production. In order to disentangle the CNM effects from the hot nuclear matter effects, $\Upsilon$ production has been studied in p--Pb collisions, in which the QGP is traditionally not expected to be formed. In pp collisions, the quarkonium production can be described as the creation of a heavy-quark pair ($q\bar{q}$) (perturbative process) followed by its hadronization into a bound state (non-perturbative process). None of the existing models fully describe the quarkonium production in pp collisions and more differential measurements can further constrain the quarkonium production models in elementary hadronic collisions.\\


 \vspace{-0.9cm}
\section{Analysis and results}
 \vspace{-0.1cm}
The ALICE Collaboration has studied bottomonium production in various collision systems (pp, p--Pb, Pb--Pb) at different center-of-mass energies per nucleon pair $\sqrt{s_{\rm NN}}$ down to zero transverse momentum ($p_{\rm T}$) and at forward rapidity ($2.5<y<4$) with the Muon Spectrometer~\cite{MCH} through the dimuon decay channel.\\

\begin{figure}[htb]
\begin{center}
\includegraphics[width=0.45\textwidth]{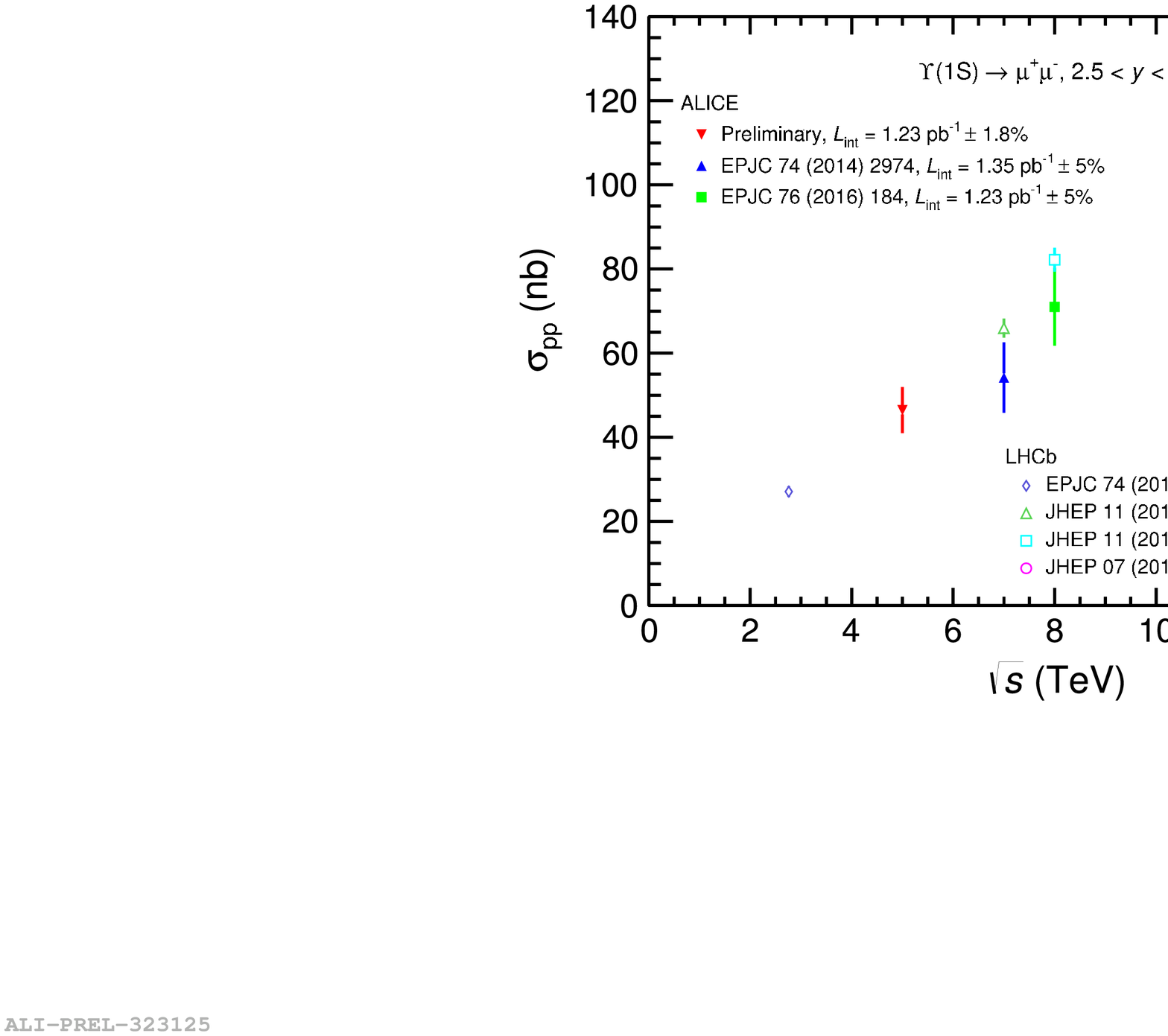}~~~
\includegraphics[width=0.45\textwidth]{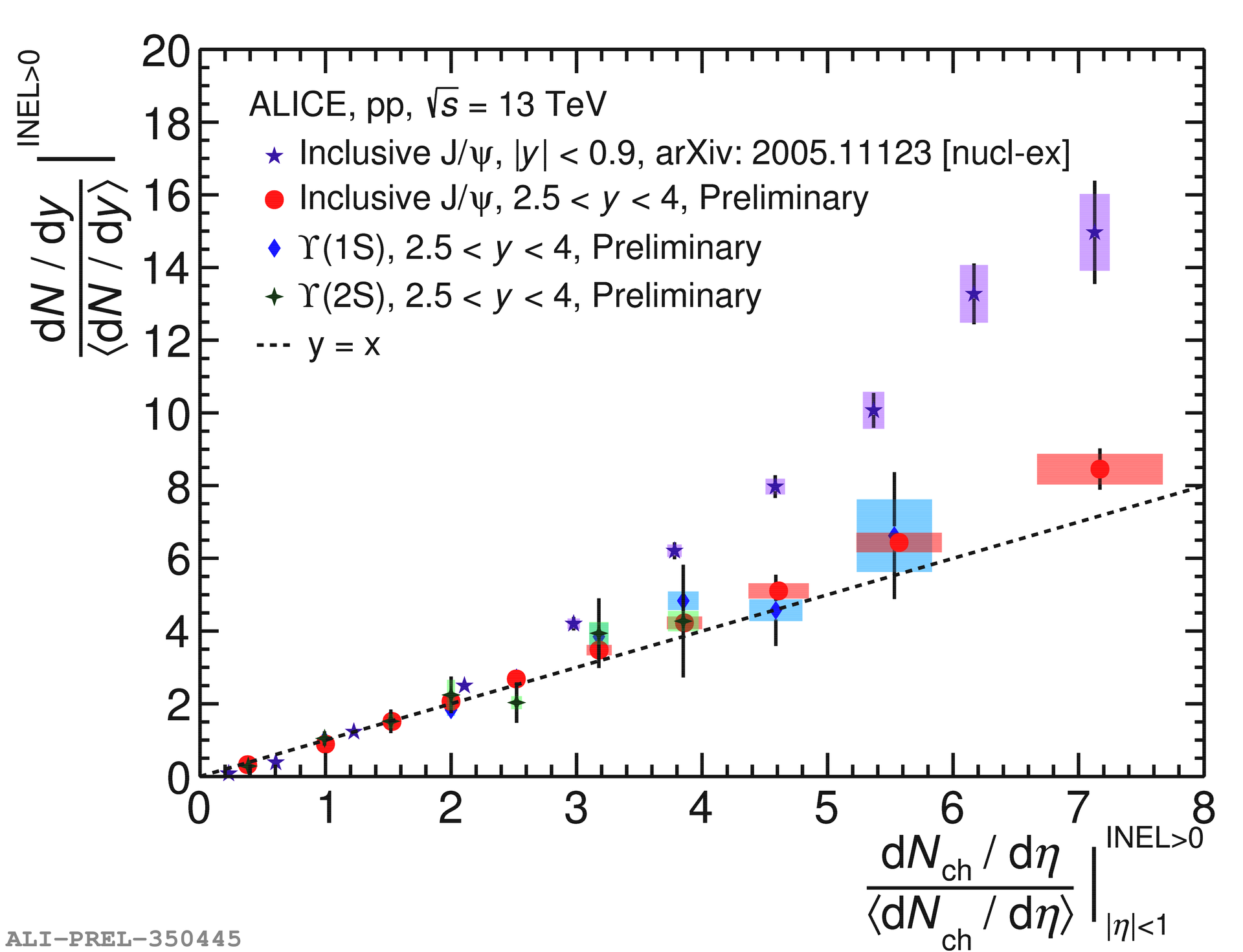}
\caption{Inclusive $\Upsilon$(nS) production cross sections as a function of the collision energy in pp collisions of ALICE and LHCb measurements (left). Relative quarkonium yield as a function of the relative charged-particle density in pp collisions at $\sqrt{s}$ = 13 TeV (right).  }
\label{fig1}
\end{center}
\end{figure}

 \hspace{-0.6cm} {\bf pp collisions}: 
The inclusive $\Upsilon$(nS) production cross sections have been measured for the first time in pp
collisions at $\sqrt{s}$ = 5.02 TeV at forward rapidity. In Fig.~\ref{fig1} (left), the energy dependence of $\Upsilon$(nS) states is shown and a steady
increase of the cross sections is observed with increasing $\sqrt{s}$. The differential cross sections as function of rapidity and $p_{\rm T}$ at $\sqrt{s}$ = 5.02 TeV have been also measured. In Fig.~\ref{fig1} (right), the bottomonium production shows a linear increase with charged-particle multiplicity at forward rapidity.

\begin{figure}[htb]
\begin{center}
\includegraphics[width=0.45\textwidth]{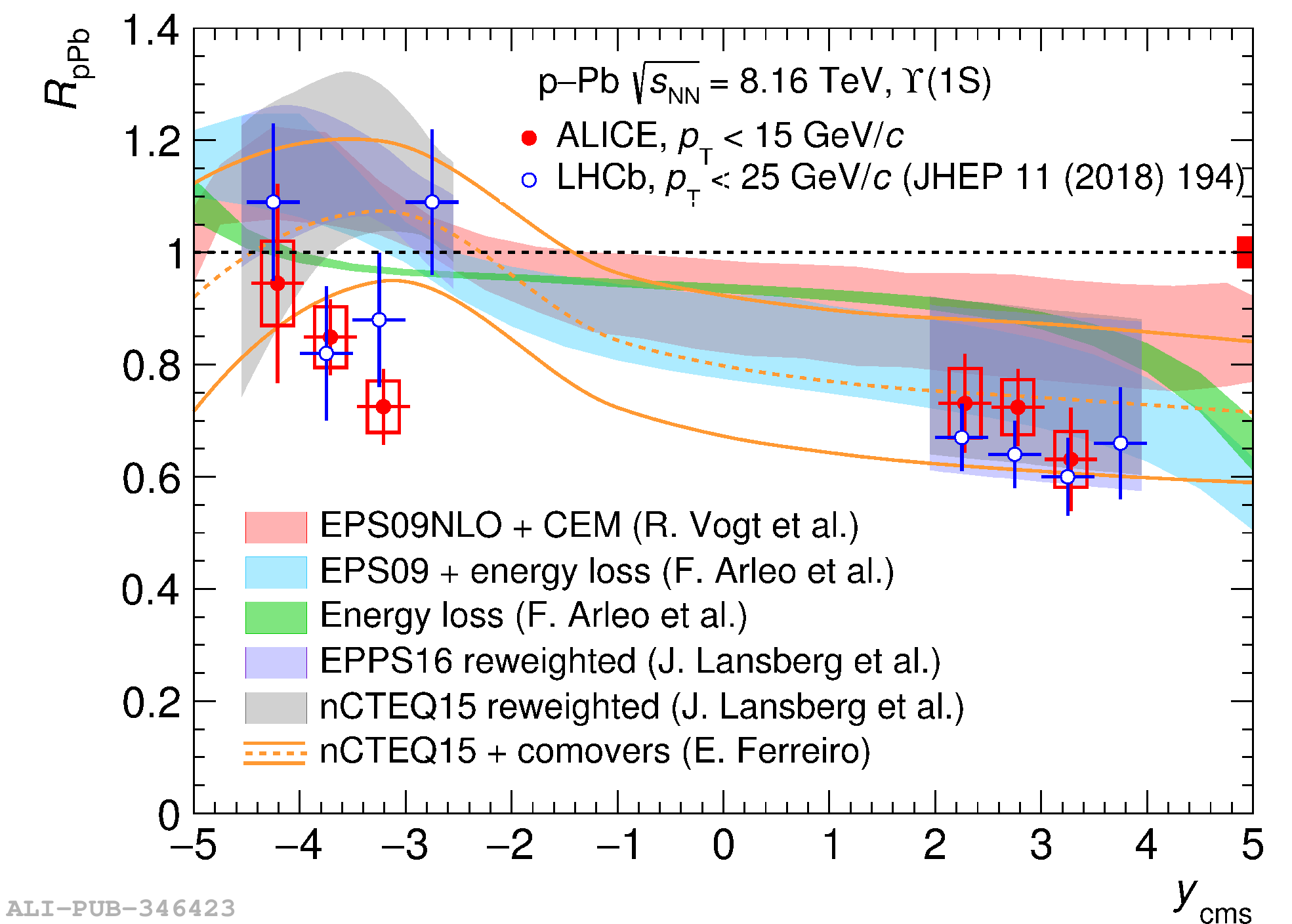}~~~
\includegraphics[width=0.45\textwidth]{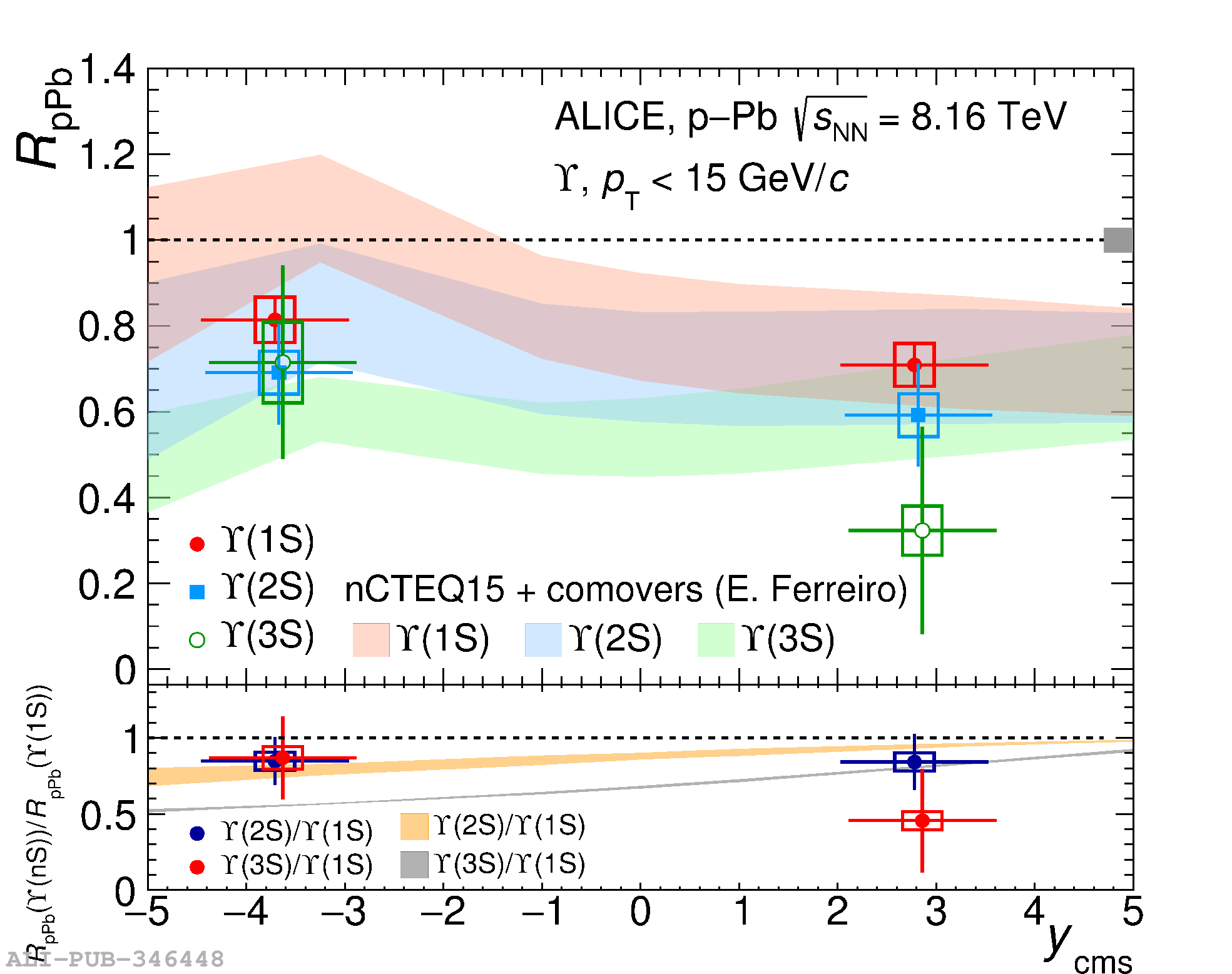}
\caption{ The $\Upsilon$(1S) $R_{\rm {pPb}} $ as a function of $y_{\rm cms}$ with different model predictions in p--Pb collisions at $\sqrt{s_{\rm NN}}$ =  8.16 TeV (left). $\Upsilon$(nS) $R_{\rm {pPb}} $ as a function of $y_{\rm cms}$ with nCTEQ15 shadowing predistion including the comover interaction.  }
\label{fig2}
\end{center}
\end{figure}

\hspace{-0.6cm} {\bf p--Pb collisions}: The CNM effects can be studied in p--Pb collisions via the nuclear modification factor ($R_{\rm pA}$), defined as \\
\vspace{-0.1cm}
$$ R_{\rm pPb} = \frac{\sigma_{\rm pPb}}{A_{\rm Pb}~.~\sigma_{\rm pp}},$$
\vspace{-0.2cm}

where $\sigma_{\rm pPb}$ and $\sigma_{\rm pp}$ are the production cross sections in p--Pb and pp collisions, respectively. $A_{\rm Pb}$ is the atomic mass number (208) of the Pb nucleus. The $\Upsilon$ production as function of rapidity, transverse momentum and multiplicity in p--Pb collisions at $\sqrt{s_{\rm NN}}$ = 8.16 TeV have been measured by ALICE~\cite{Upsi8.16ALICE}. The results show a suppression of the $\Upsilon$(1S) yields, with respect to the ones measured in pp collisions. The $R_{\rm pPb}$ values are similar at forward and backward rapidity with a slightly stronger suppression at low $p_{\rm T}$, while in both rapidity intervals there is no evidence for a centrality dependence~\cite{Upsi8.16ALICE}.
Models based on nuclear shadowing, coherent parton energy loss or interactions with comoving particles fairly describe the data at forward rapidity, while they tend to overestimate the $R_{\rm {pPb}}$ at backward-$y_{\rm {cms}}$ as shown in Fig.~\ref{fig2} (left). 
 \vspace{-0.9cm}

$\Upsilon$(2S) and $\Upsilon$(3S) $R_{\rm {pPb}}$ as a function of $y_{\rm cms}$ are shown in Fig.~\ref{fig2} (right).  The $\Upsilon$(2S) measurement also shows a suppression, similar to the one measured for the $\Upsilon$(1S), in the two investigated rapidity intervals. Finally, a first measurement of the $\Upsilon$(3S) has been performed, even if the large uncertainties prevent a detailed comparison of its behaviour in p--Pb collisions with respect to the other bottomonium states.

\begin{figure}[htb]
\begin{center}
\includegraphics[width=0.45\textwidth]{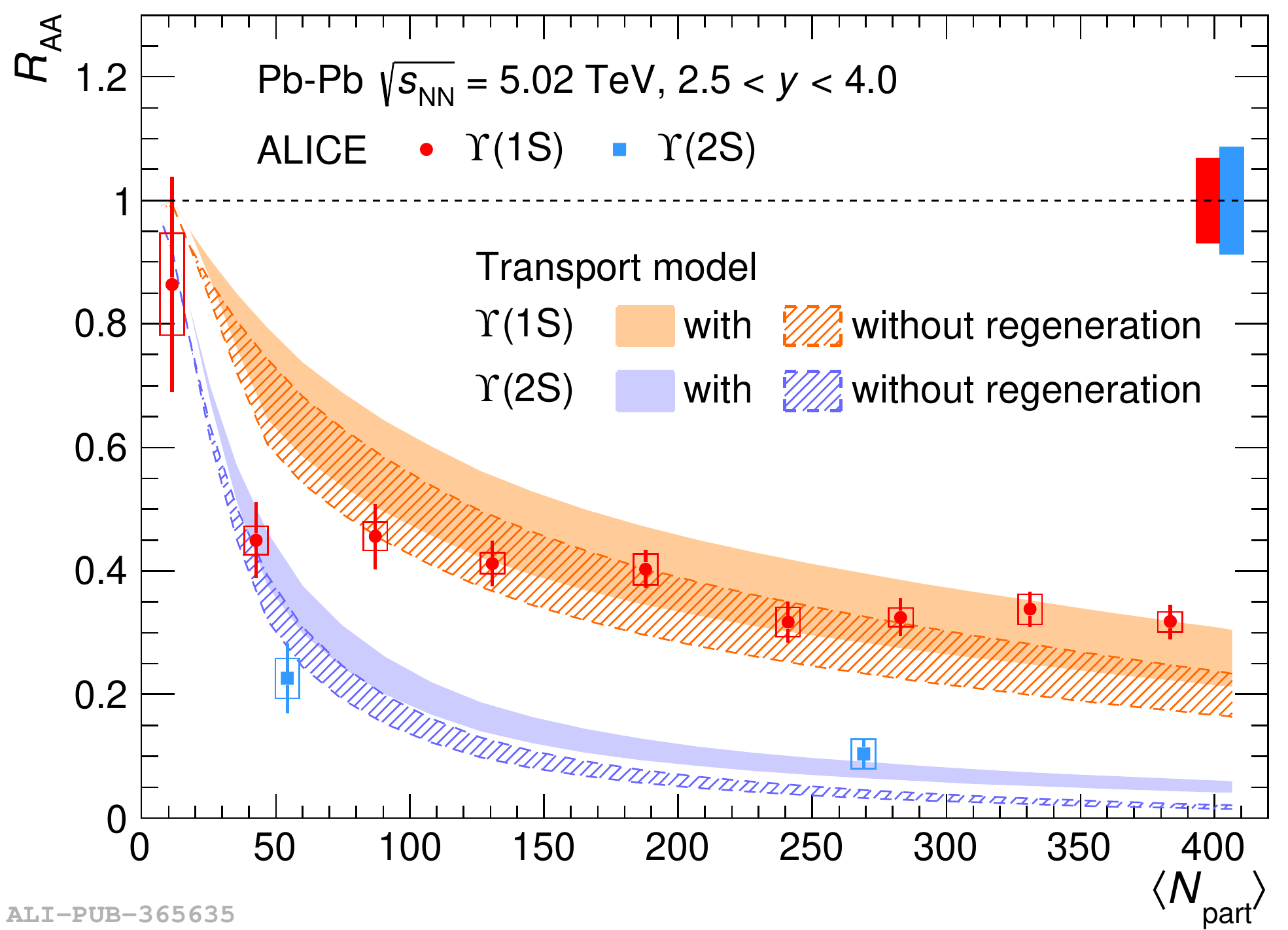}~~~
\includegraphics[width=0.45\textwidth]{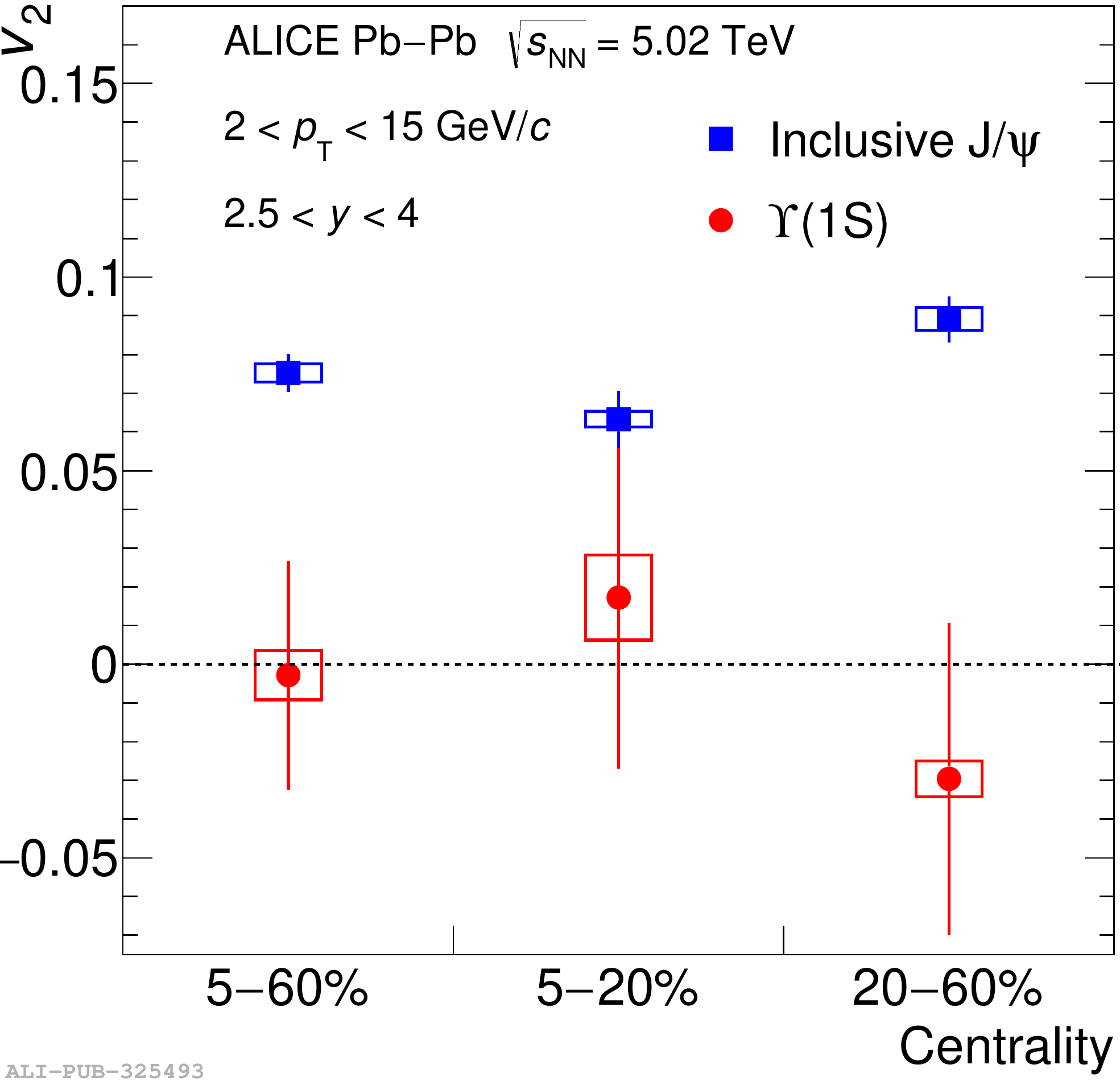}
\caption{ The $\Upsilon$(1S) $R_{\rm AA}$ in Pb--Pb at $\sqrt{s_{\rm NN}}$ = 5.02 (5.44) TeV (left). The $\Upsilon$(1S) $v_{2}$ coefficient integrated over the transverse momentum range $2<p_{\rm T}<15$ GeV/$c$ in three centrality intervals compared to that of inclusive J/$\psi$ at $\sqrt{s_{\rm NN}}$ = 5.02 TeV (right). }
\label{fig3}
\end{center}
\end{figure}

\vspace{0.2cm}
\hspace{-0.4cm} {\bf Pb--Pb collisions}: The nuclear modification factor for a given centrality class $i$ in A--A collisions can be defined as
\vspace{-0.2cm}
$$  R^{i}_{\rm AA} = \frac{{\rm d}^2N^{{\rm AA}}_{i}/{\rm d}y{\rm d}p_{\rm T}}{<T^i_{\rm {AA}}>~.~{\rm d}^2\sigma^{\rm{pp}}/{\rm d}y{\rm d}p_{\rm T}}, $$
\vspace{-0.2cm}


where ${\rm d}^2N^{{\rm AA}}_{i}/{\rm d}y{\rm d}p_{\rm T}$ is the yield in nucleus-nucleus collisions, $<T^i_{\rm{AA}}>$ is the nuclear overlap function and ${\rm d}^2\sigma^{\rm{pp}}/{\rm d}y{\rm d}p_{\rm T}$ is the production cross section in pp collisions. The measurements of $\Upsilon$(1S) and $\Upsilon$(2S) production at forward rapidity have been obtained for Pb--Pb collisions at $\sqrt{s_{\rm NN}}$ = 5.02 TeV by combining the 2015 and 2018 data sets~\cite{Upsi_PbPb5.02}. The suppression of $\Upsilon$(1S) gets stronger for the more central collisions, as shown in Fig.~\ref{fig3} (left). The rapidity dependence of $\Upsilon$(1S) $R_{\rm {AA}}$ hints at a decrease for the most forward rapidity interval~\cite{Upsi_PbPb5.02}. No significant $p_{\rm T}$ dependence of $R_{\rm {AA}}$ is observed~\cite{PbPbUpsilon,Upsi_PbPb5.02}. It is worth noting that the nuclear modification factor in p--Pb collision shows a significant $p_{\rm T}$ dependence in both forward and backward rapidity interval measured by ALICE. The different behavior between p--Pb and Pb--Pb collisions may impose some constraints on theoretical models
in near future. A larger suppression of $\Upsilon$(2S) compared to $\Upsilon$(1S) is also observed.

  \vspace{-0.6cm} 

 \vspace{-0.9cm}

 The azimuthal anisotropic ``elliptic'' flow is usually quantified in terms of the second harmonic coefficient ($v_{2}$) of the Fourier decomposition of the azimuthal particle distribution. In Fig.~\ref{fig3} (right), the $v_{2}$ coefficient of $\Upsilon$(1S) in three centrality intervals is shown, together with that of the J/$\psi$~\cite{PbPbJpsiv2}. The measured $\Upsilon$(1S) $v_{2}$ coefficient is compatible with zero within current uncertainties and this result contrasts with the J/$\psi$ $v_{2}$ measurement in Pb--Pb collisions.



\vspace{-0.2cm}

\begin{thebibliography}{8}
\vspace{-0.2cm}


\bibitem{Matsui} T. ~Matsui and H.~Satz, Phys.\ Lett.\ B 178 (1986) 416




\bibitem{regenaration1} 
  P.~Braun-Munzinger and J.~Stachel,
  Phys.\ Lett.\ B {\bf 490}, 196 (2000)
\bibitem{regenaration2} 
  R.~L.~Thews, M.~Schroedter and J.~Rafelski,
  Phys.\ Rev.\ C {\bf 63}, 054905 (2001)
  

  
  

  
  
  
  
  
\bibitem{Rothkopf:2019ipj}
A.~Rothkopf,
Phys. Rept. \textbf{858} (2020), 1-117

  
  
\bibitem{MCH}
S..~Abramovitch \textit{et al.} [ALICE Collaboration],
CERN-LHCC-99-22.


  
  
\bibitem{Upsi8.16ALICE}
S.~Acharya \textit{et al.} [ALICE Collaboration],
Phys. Lett. B \textbf{806}, 135486 (2020)

\bibitem{Upsi_PbPb5.02}
S.~Acharya \textit{et al.} [ALICE Collaboration],
[arXiv:2011.05758 [nucl-ex]].

\bibitem{Grandchamp:2005yw}
L.~Grandchamp, S.~Lumpkins, D.~Sun, H.~van Hees and R.~Rapp,
Phys. Rev. C \textbf{73} (2006), 064906
  

  
  
  \bibitem{PbPbJpsiv2} 
  S.~Acharya {\it et al.} [ALICE Collaboration],
  Phys.\ Rev.\ Lett. {\bf 119}, no. 24, 242301 (2017)
  
  


  
  
\bibitem{PbPbUpsilon} 
  S.~Acharya {\it et al.} [ALICE Collaboration],
  Phys.\ Lett.\ B {\bf 790}, 89 (2019)
  

  
  
\bibitem{UpsilonV2}
S.~Acharya \textit{et al.} [ALICE Collaboration],
Phys. Rev. Lett. \textbf{123} (2019) no.19, 192301
  





 







\end{thebibliography}
\end{document}